\documentclass[11pt,a4paper]{article}

\usepackage{setspace}
\usepackage{apacite}
\usepackage{natbib}

\usepackage{lineno}

\usepackage{xr}
\usepackage{color}
\usepackage{amsmath,amsthm,amssymb,color,array,multirow}
\usepackage{epsfig,amsfonts,latexsym, hyperref,  mathrsfs}
\usepackage{tikz}
\usetikzlibrary{shadows,trees}
\usepackage{amscd}
\usepackage{fancyhdr,latexsym,amsmath,amsfonts,amssymb,amsbsy,amsthm,url}
\usepackage[hscale=0.7,centering]{geometry}
\usepackage{graphicx,epsfig, xy}
\usepackage{hyperref}
\usepackage{lscape,fancybox}
\usepackage{color}
\usepackage{caption}
\usepackage{subcaption}
\usepackage{comment}

\usepackage{algorithm}
\usepackage{algcompatible}

\newcommand{\blmma}{\begin{lemma}}
\newcommand{\elmma}{\end{lemma}}

\newtheorem{lemma}{Lemma}

\title{\textbf{Eliciting core spatial association from spatial time series: a random matrix approach}}
\date{
Madhuchhanda Bhattacharjee, University of Manchester\footnote{email: chhanda.bhatta@gmail.com}
\\
Arup Bose, Indian Statistical Institute, Kolkata
\\
April 2026
}

\begin{document}

\def\shortauthors{ABC}
\maketitle


\normalsize
\begin{abstract}
\begin{singlespace}
Spatial time series (STS) data are fundamental to climate science, yet conventional approaches often conflate temporal co‑evolution with genuine spatial dependence, obscuring subtle but critical climatic anomalies. We introduce a Random Matrix Theory (RMT)–based framework to isolate ``core spatial association'' by suitably trimming out strong but routine temporal signals while preserving spatial signals. 
Our pipeline introduces Hilbert space filling curve technique and Bergsma’s correlation measure of statistical dependence, to climate modelling. Applied to the diurnal temperature range (DTR) data of India (1951–2022), the method reveals distinct spatial anomalies shaped by topography, mesoclimate, and urbanization. The approach uncovers temporal evolution in spatial dependence and demonstrates how regional climate variability is structured by both physical geography and anthropogenic influences. Beyond the Indian application, the framework is broadly applicable to diverse spatio‑temporal datasets, offering a robust statistical foundation for predictive modelling, resilience planning, and policy design in the context of accelerating climate change.  
\end{singlespace}
\end{abstract}

\maketitle

\newpage
\section{Introduction} 

Understanding spatial association in climate and environmental data is central to advancing predictive modelling, risk assessment, and resilience planning. Spatial time series (STS)—datasets that capture both temporal evolution and spatial variability—are increasingly used in studying phenomena ranging from temperature anomalies and precipitation extremes to multi‑hazard interactions. Yet, extracting the core spatial association embedded in such data remains a methodological challenge. 

Regrettably, spatial association is an ambiguous term in spatial literature. In majority of cases, it means whether the values of a variable are determined/influenced by the location. 
However, this feature is self-evident for climate variables, and can be illustrated by a simple map. We set out to elicit critical understanding of how for a climate variable, change at a spatial location impacts its value at other locations, by an easily transferable model that does not depend on climate or region specific parametrisations. 

Spatial time series (STS) data in climate science are dominated by strong temporal co‑evolution, which often obscures subtle or even well‑known spatial anomalies. Conventional spatial association measures, when applied directly to raw time series, fail to disentangle genuine spatial dependence from temporal dynamics, particularly in high‑dimensional, non‑stationary, and strongly coupled systems. Standard detrending methods either strip away essential spatial features or prove ineffective, while model‑based approaches that assume separability of temporal and spatial effects are rarely realistic. To uncover core spatial behaviour, temporal influences must be carefully dampened without erasing meaningful spatial signals.


We propose a novel de-trending pipeline for STS data using Random Matrix Theory (RMT). This process successfully eliminates the dominant influence of time while at the same time able to reveal significant spatial features. RMT offers a powerful statistical framework for addressing the current challenge. Originally developed in physics to study complex spectra, RMT has since been applied to finance, neuroscience, and network science, where it provides robust tools for distinguishing signal from noise in large correlation matrices. Its application to spatial time series in climate science is both novel and timely. By leveraging singular-value spectra and universality properties, RMT enables the identification of statistically significant signals, thereby isolating the temporal and spatial  structures that govern system dynamics.

As mentioned earlier, spatial association is often conveyed through maps that display the distribution of variables of interest, using color patterns to illustrate variation across locations. While such visualizations provide an intuitive understanding of spatial heterogeneity, they are limited from a modeling perspective. Crucially, they do not quantify the statistical relationship between two locations for the same variable. This distinction is particularly important in the current climate context, where understanding how changes in one location influence conditions in another is essential for anticipating cascading impacts and designing effective resilience strategies.

Towards numerically capturing spatial association from spatial time series (STS), the literature typically employs measures such as Moran's $I$, Geary’s $C$, and related indices. These approaches rely on two matrices: one encoding the proximity of spatial locations, and the other representing the “similarity” between values of a variable across pairs of locations. Numerous similarity measures have been proposed, most of which rest on the assumption of repeated, independent, and identically distributed (i.i.d.) observations for each location pair. However, the presence of strong temporal patterns in STS inevitably violates the i.i.d. assumption, rendering such implementations problematic and potentially misleading. Addressing this limitation is essential for developing robust methods that can faithfully capture spatial association in the context of temporally structured climate and environmental data.

Once we have a properly time‑detrended data, the assumption of independence becomes approximately valid, enabling us to explore spatial association in several novel ways. First we consider the high-dimensional correlation matrices and denoise it using the Mar\v{c}enko-Pastur law (from RMT) which is based on eigenvalue spectra and universality properties. For graphical presentation of these correlation matrices special care should be taken to ensure that the spatial proximity is preserved in such linear/one-dimensional presentation of two-dimensional locations. This we achieve by arranging them following Hilbert-space-filling-curve technique.

For smaller dimensional summary of these rather large matrices, we use empirical spectral distributions (ESD), which are highly useful for comparing and contrasting correlation matrices arising under different situations.  Finally, we use global measures as univariate summaries of association. We  study statistical dependence between the locations using advanced measures like Bergsma's correlation, apart from linear association measures, such as Pearson's correlation, and for both we investigate the global spatial association using various spatial weight matrices.

We demonstrate  the method through the analysis of the diurnal temperature range (DTR) data of India. Our method is easily implementable for other types of spatio-temporal (climatic) data. DTR data have been investigated at the global level (\cite{stjern2020} and \cite{zhong2023}), and also at regional levels. See \cite{kothawale2010},
\cite{jhajharia2011},
\cite{vinnarasi2017},
\cite{roy2019},
\cite{sharma2021},
\cite{mall2021} and \cite{jayasankar_misra2024} for work on data related to the Indian sub-continent. These works have primarily focussed on the temporal pattern. Some have recognized the existence of spatial variations that are masked by strong temporal components. The \textit{Multidimensional Ensemble Empirical Mode Decomposition} (MEEMD) developed by \cite{wu2009} 
has been applied by \cite{ji2014} and \cite{vinnarasi2017} to climate data to eliminate the oscillatory component of a time series and reveal the slow varying components. However, these techniques do not account for dependence between DTR series from different geographic locations.
Data sets from other geographical regions are expected to exhibit their own intrinsic attributes, and as an illustration, we briefly look at the daily DTR data from Bahia, Brazil. 

It is to be noted that our method is actually not restricted to spatio-temporal data, and can be implemented on any data which have two dimensions, with strong signals from one of the two dimensions.
This generality makes the framework a powerful tool for uncovering hidden structures across diverse scientific domains, far beyond climate applications.

\section{Materials and methods}

\subsection{Data}\label{sec:materials} 
\begin{figure}[!ht]
     \centering
     \includegraphics[scale=.5]{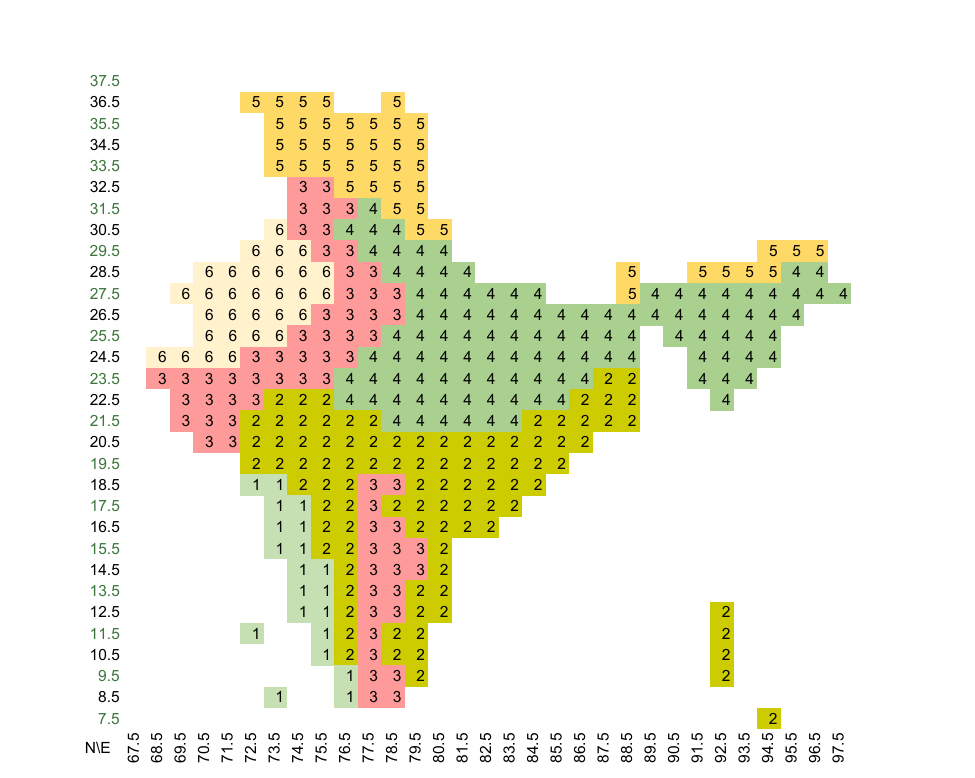}\ \ \caption{362 grids 
     of India: 1-Tropical monsoon; 2-Tropical savannah, wet and dry; 3-Arid, steppe, hot; 4-Humid subtropical; 5-Montane climate; 6-Hot deserts, Arid.}
    \label{fig:India_climate_map.pdf}
\end{figure}

India is divided into 362 grids across six climatic zones at $1^{\circ}\times 1^{\circ}$ resolution. 
The stylised gridded map of India (Figure \ref{fig:India_climate_map.pdf}) emphasises the extent of spatial coarseness of the data.  

The daily maximum and minimum temperatures in Celsius, at each of these grids, for the period 1951-2022, is available from \cite{crs}.
The difference between this daily maximum and minimum has been taken as a proxy for the DTR value for a 24 hour period for that day.

Let $X$ be the \textit{two dimensional data array} containing these values over 72 years yielding a 362-dimensional vector time series with 26298 time points, visualized as a $26298\times 362$ rectangular matrix. For a specific year
the rows are arranged in the naturally increasing order of time. For now, the 362 locations (columns) are arranged in some fixed order (often the order in which the data is available from its source). The data shall be grouped by months, seasons, or years in time and by the six climatic regions of India in space, depending on the specific aspects that we wish to probe. Some values are missing and complete data are available only for 280 of the 362 locations. Thus, when we work with the individual values, we consider the 280 grids. However, the missingness is such that if aggregated data for months or years are required, then it is possible to use all 362 grids and adjust for the number of missing values while taking the averages.

\subsection{Methods}\label{sec:methods} 


We begin with an itemized outline of our approach, elaborated in detail later:
(i) We first reorder the columns of $X$ to improve visual interpretation of results, yielding the rearranged matrix $D$.
(ii) Using time‑series techniques, we obtain a detrended matrix $T$ from $D$, though this also removes some spatial signals. 
(iii) From the empirical distribution of singular values of $D$, we construct a trimmed matrix $S$, and confirm via generalized singular value decomposition (GSVD) that $S$ is a reliable proxy for $D$. This lower‑rank matrix dampens temporal effects while retaining spatial signals.  
(iv) With $S$ established, we compute the Pearson correlation matrix to uncover novel patterns of core spatial association.  
(v) The statistical dependence between the grids is studied by Bergsma's correlation measure and global spatial association is quantified using various spatial weight matrices. 
(vi) All these quantities are assessed at different levels of spatial resolution and temporal windows. Also we used different mathematical abstractions, like studying the entire correlation matrix  or studying it's ESD, etc.
\vskip5pt

We now explain the above ideas in more detail. 

\noindent \textbf{(i) A novel linear ordering of spatial locations:} Each time series of a grid has two additional pieces of information, namely their latitudes and longitudes. Therefore, to investigate the temporal and/or spatial behavior of the DTR across times and regions, we need to first decide how this two-dimensional location information should be incorporated so as to arrive at an appropriate ordering among these time series and hence an appropriate data matrix for the entire DTR data. 

We discover an appropriate ordering by trial and error, and this ordering leads to an excellent visual display of spatial association, as seen later in various figures in the paper.
We first stratify the grids according to their climatic zones. Then we employ the Hilbert space-filling curve arrangement of the locations, for example, as follows:
we start from the lowest latitude and longitude position, say position $(1,1)$. Then we move as follows:
$(1, 1)\rightarrow (1, 2)\rightarrow (2, 1)\rightarrow (3, 1)\rightarrow (2, 2)\rightarrow (1, 3)$ and so on. 
In other words, we move anti-diagonally and wrap around. 
We refer to this as the \textit{spiral ordering}, 
which provides a \textit{linearized ordering} of the two-dimensional spatial locations. 
Thus we now have a 
$26298\times 280$ \textit{reorganized data} matrix $D$ from $X$.


For non-gridded data (e.g. the Bahia-Brazil data illustrated later), the same principle still works, but it requires some coding to disentangle the in-homogeneously placed two-dimensional spatial points to arrive at a reasonable linear ordering.
\vskip5pt

\noindent
\textbf{(ii) Traditional time detrending:} We employ a traditional time detrending method, such as time series decomposition, on every time series at a location. This yields the matrix $T$ from $D$. Then we first obtain the (Pearson) correlation matrix $R^T$ for  $T$, followed by its eigenvalues. Next, we apply the Mar\v{c}enko-Pastur law on the eigenvalues of $R^T$ to evaluate the degree of significant signals contained in them. 
\vskip5pt

\noindent \textbf{(iii) SVD based dimension trimming of the data matrix:} 
We then obtain a low-dimensional \textit{trimmed data matrix} $S$ from $D$, using SVD-based approximation. This is necessary to remove the dominant temporal trends, but has to be done without losing any significant amount of information. First, the number of significant singular values, say $s$, of $D$ is identified, based on the null distribution of singular values obtained by empirical methods, such as permutations. Then we successively \textit{remove} contribution of the top significant singular values until the \textit{autocorrelation function} (ACF) of the resulting trimmed matrix is within a target threshold, or all significant singular values have been exhausted, whichever is earlier. We arrive at the trimmed matrix as
$$ S = D - \sum_{k=1}^{d} \lambda_{k}  u_{k} v^{T}_{k},$$
where 
$p$ (for us $p=280$) is the dimension of $D^TD$, 
$d < s \leq p$, and $\lambda_j$, $u_j$, $v_j$ are the $j$th singular value,  left singular vector and right singular vector, respectively, for $j = 1, \ldots, p$. 
Note that the top singular values are removed rather than retained, so that we get rid of the significant time components. 
The extent of time detrending in $S$ is assessed by the ACF, as mentioned above.
The retention of spatial signals in $S$ is discussed next.
\vskip5pt

\noindent \textbf{Assessing information retention using GSVD:} The generalized singular value decomposition (GSVD) 
simultaneously 
decomposes a pair of matrices that have the same number of columns. This has been used in bioinformatics for comparing two datasets, such as for finding common gene expression patterns in two different studies, even if they have different (sets of) individuals. 
We have used GSVD to confirm that the move from $D$ to $S$ does not entail any significant distortion of the spatial information. Thus, the lower dimensional $S$ can be used as a surrogate for $D$ to explore spatial association. 
Since no relevant random matrix results are known for the distribution of GSV in the null case, the observed GSV distribution will be compared with the critical values of an empirically obtained (null) distribution of GSVs. 
\vskip5pt


\noindent
\textbf{(iv) Core spatial association:} Once the reduction to $S$ has been made, Pearson correlation matrix $R^{S}$ of $S$ could be used as a measure of Core Spatial Association. Multiple further analyses can be performed based on $R^{S}$. First, we may look to remove any noise from $S$ using the Mar\v{c}enko-Pastur(MP) law. Following which, we may present either the original or the MP-denoised matrix  in a graphical form. We could investigate if there has been any temporal variations in the spatial association and obtain, e.g. yearly spatial association matrices. However, this leads to the issue of comparing multiple high-dimensional correlation matrices. We  use RMT based technique of 
ESD
for these comparisons, 
thereby reducing the problem. In addition, 
we could also use the MP law to study the number of significant signals in $R^{S}$. 
It goes without saying that global measures like Moran's $I$, based on the individual Pearson correlations within $R^{S}$, 
 can also be used, to arrive at a univariate summary of association.

\vskip5pt

\noindent
\textbf{(v) Spatial Bergsma--alternative to Pearson's correlation:} 
However, climate data is typically 
non-Gaussian, 
and exhibits non-linear dependence. Hence 
the use of Pearson's correlation is perhaps not appropriate.
We use a non-parametric measure, namely Bergsma's 
correlation $\rho$, which captures statistical dependence and not merely linear association. 
See \cite{bose} for details on the estimation of 
$\rho$. 
This leads to the \textit{spatial association measure} $S_B$ (\textit{Spatial Bergsma}) of \cite{kappara}
as follows. Suppose that we have $p$ spatial units (for us, $p=280$), and let $X_i$ denote a real variable observed at the location $i$, $1\leq i\leq p$. Then $S_B$ with a row-standardized $W$ is defined as 
\begin{equation}
S_{B}
 = p^{-1}\sum_{1\leq i <j \leq p}(w_{ij}+w_{ji})\rho^{(ij)}.\label{SBp1}
\end{equation}
Here, $\rho^{(ij)}$ is Bergsma's correlation between $X_i$ and $X_j$. The entries $w_{ij}$ are numerical weights assigned to $(i,j)$, and larger weights signify greater spatial proximity. It is always assumed that $w_{ii}=0$ for all $i$. See \cite{getis2010analysis} for popular choices for $W$. We shall use two choices. The \textit{lag-1 adjacency matrix} uses $w_{ij}=1$ if $i$ and $j$ are (geographically) adjacent, and $0$ otherwise. For a second choice, we use exponential decay on Euclidean distances between the grids. The parameter $S_{B}$ is estimated by the spatial Bergsma statistic,  
\begin{equation}
	\tilde{S}_{B}:=p^{-1}\displaystyle{\sum_{1\leq i <j \leq p}(w_{ij}+w_{ji})\tilde{\rho}^{(ij)}}\,,\label{sb}
\end{equation} 
where $\{\tilde{\rho}^{(ij)}\}$ is an estimate of $\rho^{(ij)}$. 
Calculation of $\rho^{(ij)}$, and hence of $\tilde{S}_{B}$ is computationally intensive. The {\tt R} code of \cite{bose} may be used for this calculation. 
The asymptotic distribution of $\tilde S_B$ is known 
when the observations over time are i.i.d. We use it as a comparative and diagnostic tool after time-detrending the data.
Apart from using this global measure, the correlation matrices, consisting of pairwise Bergsma correlations (denoted by $B^{S}$), can also be used to study dependence in detail.
\vskip5pt

\noindent \textbf{(vi) Temporal changes in spatial association matrix:} 
The correlation matrices, whether Bergsma's or Pearson's, have been computed for various temporal windows and/or spatial regions, e.g.  
for the entire period or for individual years, for each month of each year, etc., and for the entire region or for individual climatic regions, etc. 
These correlation matrices 
can be summarized by their ESDs,
or by a univariate global measure, 
and then compared to assess temporal and regional changes in the core spatial association.  
\vskip5pt

\noindent
\textbf{Temporal changes in spatial Bergsma:} Spatial Bergsma enables us not only to address 
non-Gaussianity, but also to 
summarise each correlation matrix
into a single numerical value (similar to global Moran's $I$ statistic). Thus, it provides easy graphical summary to capture changes in spatial association over years (or months), at the aggregate level or at the climatic region level.
\vskip5pt

\textbf{Algorithm 1} shows the work flow of our method.  
%


\begin{algorithm}
    \caption{Core Spatial Association}\label{csa}
    \begin{algorithmic}
        \STATE  $X$: Spatial time series data from $p$ locations         
        \STATE  `spiral ordering': Arrange the $p$ locations in this ordering 
        \STATE  $D$: Resulting data matrix, which is the column rearranged $X$ matrix \\

        \STATE  SVD: Carry out Singular Value Decomposition of $D$
        \STATE  $\lambda_j$, $u_j$, $v_j$: The $j$-th singular value, left singular vector and right singular vector, respectively, for $j = 1, \ldots, p$ \\

        \State s: Number of significant singular values \Comment{This is a comment}
        \State $\, \, \, \, \, $   Obtained by using empirical methods like permutation \Comment{This is a comment} 
        \State $\, \, \, \, \, $   Then the following loop should be j = $1$ to $s$ \Comment{This is a comment} \\
        \FOR{j = $1$ to $p$}
            \STATE Calculate $S = D - \sum_{k=1}^{j} \lambda_{k}  u_{k} v^{T}_{k}$
            \STATE Obtain Autocorrelation Functions
            \IF {ACF within threshold} 
                \STATE d = j
                \STATE break forloop
            \ENDIF
        \ENDFOR 
        \State d: Minimum number of singular values to trim D  \Comment{This is a comment}
        \State $\, \, \, \, \, $   Obtained by checking if ACF is within threshold.  \Comment{This is a comment} \\

        \STATE $S = D - \sum_{j=1}^{d} \lambda_{j}  u_{j} v^{T}_{j}$
        \STATE $S$ is the trimmed data  \Comment{This is a comment} \\
        
        \STATE $R^{S}$ = Pearson Correlation matrix of S 
        \STATE This is used as Core Spatial Association  \Comment{This is a comment} \\
        
        \STATE Further analyses on $R^{S}$   \Comment{The followings are comments}
        \STATE (1) Study association with respect to a fixed location
        \STATE (2) Use Marchenko-Pastur law to study/use the number of significant signals in $R^{S}$
        \STATE (3) ESD of overall $R^{S}$ or $R^{S}_{i}$ where $S$ is restricted to data from the $i$-th year\\
        
        \STATE Further analyses with Spatial Bergsma ($S_B$)   \Comment{The followings are comments}
        \STATE (4) Calculate global $S_B$ statistic using $B^S$, Bergsma's correlation, and spatial weight matrices $W_1$ or $W_2$ 
        \STATE (5) Obtain series of $S_B$ statistics at yearly (or monthly) level using $B^{S}_{i}$
        \STATE (6) Study $S_B$ statistics for each climatological region 
    \end{algorithmic}
\end{algorithm}
\normalsize


\section{Results and discussion}

\subsection{Raw spatial association pattern}\label{sec:raw_spatial_analysis} 

\textit{Pearson's correlation matrices}: 
Applying the MP law-based cutoff, of the 280 eigenvalues, only the top $10$ eigenvalues $\{\lambda^{D}_i\}$ (with the eigenvectors $\{e^{D}_i\}$) of $R^D$ are significant for the original data matrix $D$. Also the largest eigenvalue is in magnitude ten times the second largest. This may not be surprising, since common temporal patterns arising due to seasonality would naturally lead to a strong association/correlation. However, the presence of a high degree of positive association might be masking other patterns. 
Thus, $R^{D}$ is approximated by the MP law based \textit{de-noised matrix} $\hat{R}^{D}$ : 
\[\hat{R}^{D} = \sum_{j=1}^{10} \lambda^{D}_{j}  e^{D}_{j} {(e^{D}_{j})}^{\top}.
\]
    \begin{figure}[ht!]
    \centering
    \includegraphics[width=0.35\linewidth]{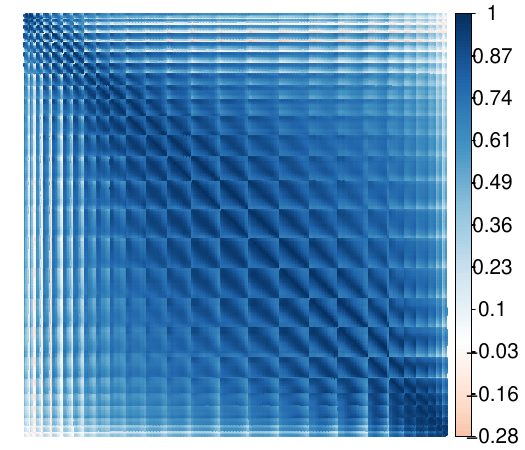}
    \caption{\textit{Upper triangle}--entries from Pearson correlation matrix ($R^D$) based on original data (D); \textit{Lower triangle}--from MP de-noised $R^D$.}
    \label{fig:DTR_cor_MPdns}
\end{figure}
The upper and lower triangles in Figure \ref{fig:DTR_cor_MPdns} are based on $R^D$ and $\hat{R}^{D}$ respectively. 
MP law based de-noising has some effect and brings balance to the correlation distribution (observe the negative values along the left vertical). However, the distribution is still highly positive, which is not surprising given the observation made above regarding the top eigenvalue. Thus, although the MP law has been very useful in many applications, it is not effective here in bringing out any pattern, when implemented on $R^{D}$. 

In addition to the full correlation matrices, we also studied the correlation matrices based on data for every year $i$. Over the years there has been a small but steady increase in the median of the ESDs, and a slight shift away from zero in the left tail of the distribution. Since the sum of eigenvalues is fixed for a correlation matrix, the above, generally speaking, will mean that the right tail too has shrunk towards the center. This could be indicative that multiple signals are determining the system or that multiple factors are influencing the system. 


\subsection{Elucidating core spatial association}\label{sec:core_spatial_analysis} 

\subsubsection{Temporal trend adjusted spatial signals}\label{sec:spatial_signals} 

\noindent\textbf{Traditional time detrending}: Figure \ref{fig:DTR_cor_MPdns} indicates that to bring out spatial association patterns, it is imperative to remove the association due to predominant temporal co-evolution. We first followed standard detrending technique and eliminated the linear trends along with seasonal components from each of the 280 time series. The correlation matrix $R^{T}$ for the so adjusted data $T$ has mostly positive entries, and large chunks of near zero entries. By applying the MP law on $R^{T}$, we find only $18$ significant eigenvalues (out of 280), with only one dominant eigenvalue. In view of the common knowledge of
multiple topographic, orographic and other sources of spatial climatic anomalies in India, we can conclude that the standard time detrending technique has failed. 
\vskip5pt

\noindent\textbf{Significant singular values:} A 
standard method to understand the signal or information content of a matrix is through its singular values. Large values correspond to major information content, while small ones represent less significant features. 
Since $D$ 
can be viewed 
as a  \textit{random} matrix,
we need to also keep in mind the randomness of its entries.
The distribution for the singular values based on the DTR data was obtained empirically, using the permutation method and 500 times permuted DTR data. According to the empirical distribution, there were only 3 singular values that were not significant. 
\vskip5pt

\noindent \textbf{A novel SVD based time detrending:} Singular value decomposition (SVD) was used to remove dominant temporal patterns and obtain an approximation of $D$. The efficacy of this detrending will be assessed by using the values of the ACF, and the overall signal coverage as measured by the \textit{percent share of SVs} retained. We identified the minimal number of (significant) SVs of $D$ required to reasonably \textit{time-detrend} the data, while retaining as much of the overall signal as possible.  
\vskip5pt
\begin{figure}
\centering
\begin{subfigure}{.4\textwidth}
  \centering
   \includegraphics[width=0.8\linewidth]{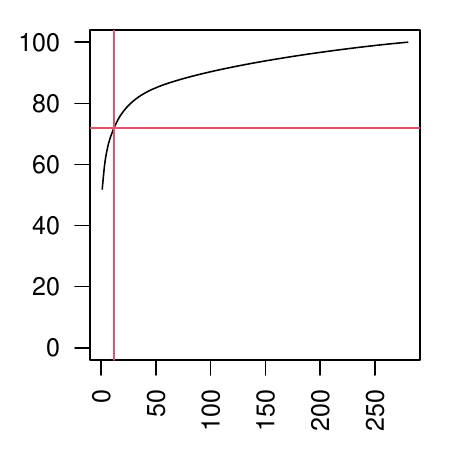}
    \caption{}
    \label{fig:Singular_values_of_org_DTR}
\end{subfigure}
\begin{subfigure}{.4\textwidth}
  \centering
    \includegraphics[width=0.8\linewidth]{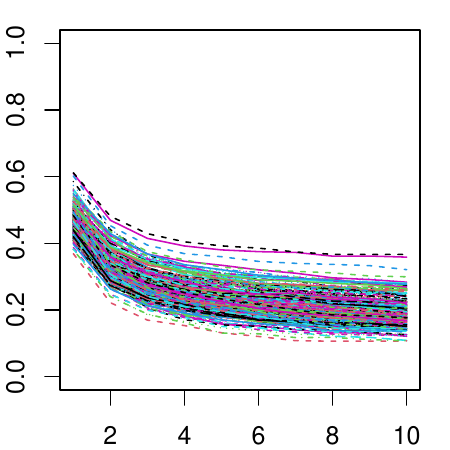}
    \caption{}
    \label{fig:ACF_of_org_DTR_data}
\end{subfigure}
\caption{(a) Left panel: Cumulative singular-values for original DTR data ($D$). \\
(b) Right panel: Autocorrelations for the 280 time series based on trimmed data ($S$).
}
\label{fig:DTR-SVD}
\end{figure}
\noindent
For Indian DTR data, we used the above two criteria (i.e. low autocorrelation and retention of significant singular values) and removed
the top 12 SVs (and vectors), to obtain the adjusted
\textit{core spatial  data matrix} $S$. Referring to Figure \ref{fig:Singular_values_of_org_DTR}, these cover 72\% of the total sum of SVs.  
The autocorrelation plots of the time series of each column of $S$, given in Figure \ref{fig:ACF_of_org_DTR_data}, show that this trimming removed much of the temporal pattern. Applying the MP law on the ESD of $R^S$ now shows $33$ significant eigenvalues, compared to only $10$ for $R^D$ and $18$ for $R^T$, as mentioned earlier. We also consider entire ESDs and observe multiple positive effects of the proposed trimming method (see \textit{Supporting Information} document, \cite{bhatbosesupport}).



\subsubsection{Spatial climatic anomalies in India} 

Topography plays a pivotal role in shaping the spatial association of temperature across India, primarily through altitude‑induced cooling, orographic influences on precipitation, and thermally driven wind circulations in mountainous regions. In particular, the Himalayan terrain and heterogeneous urban landscapes act as key topographic modifiers of local and regional temperature regimes. These features generate distinct “cold spots” and “hot spots” that diverge markedly from surrounding areas, underscoring the importance of topographic complexity in driving spatial climate variability. 

\begin{figure}[h]
    \centering
    \includegraphics[width=0.6\linewidth]{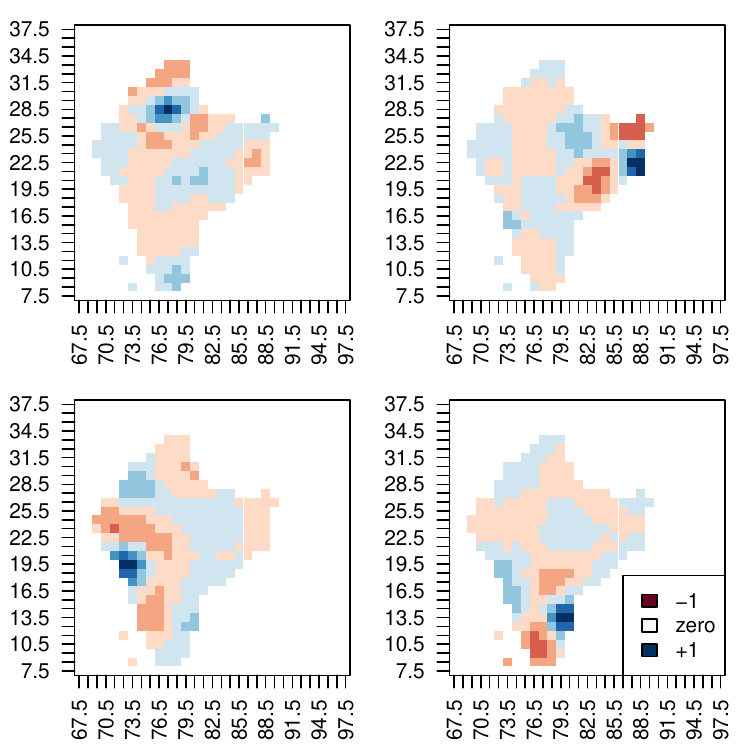}
    \caption{Correlations with respect to locations  
    of the four major cities. \textit{top left}, Delhi; \textit{top right},  Kolkata; \textit{bottom left}, Mumbai; \textit{bottom right}, Chennai, based on trimmed data.}
    \label{fig:IMD_DTR_SVD12s_res_corr_4cities.pdf}
\end{figure}

Such anomalies can be investigated using the core-spatial signal $S$ that was derived using SVD based trimming.
For instance, consider the behavior of the elements of the correlation matrix $R^S$ 
corresponding to some of the major cities, as presented in Figure \ref{fig:IMD_DTR_SVD12s_res_corr_4cities.pdf}. These heat maps exhibit strong spatial patterns. 
Some of these can be attributed to known anomalies (see below), whereas some novel patterns merit 
further investigation.

\noindent\textbf{Urban heat islands:} Large cities such as Delhi and Kolkata exhibit pronounced surface temperature anomalies, amplified by reduced cloud cover and the expansion of the Planetary Boundary Layer (PBL). In India, the PBL represents the lowest portion of the atmosphere (typically extending 1,000-2,000 m above the surface) that directly interacts with the land through exchanges of heat, moisture, and momentum. It displays a distinct diurnal cycle—convective and turbulent during the day, but stable at night. These dynamics disrupt smooth spatial temperature behaviour, often generating localized anomalies that manifest as negative correlations or weakened associations with surrounding areas.
This is clearly supported by Figure \ref{fig:IMD_DTR_SVD12s_res_corr_4cities.pdf}.
\vskip5pt

\noindent\textbf{Meso-climate:} Mesoclimate in India refers to localized climatic conditions at scales of roughly 10–100 km, shaped by terrain features such as mountains, valleys, coastlines, and urban environments. Prominent mesoclimatic zones include the humid coastal belts, the cooler Himalayan highlands, the arid and semi‑arid regions of Rajasthan, and the tropical humid zones of the northeast, each influenced by local wind systems such as land–sea breezes and mountain winds. Examples include: (1) Urban Mesoclimates—-Delhi exhibit pronounced “urban heat island” effects, creating warmer conditions relative to surrounding rural areas; and (2) Coastal Mesoclimates—-Mumbai and Chennai are characterized by warm, humid conditions with moderated temperature variability driven by sea breezes. These mesoclimatic distinctions highlight the critical role of local geography and land use in shaping India’s diverse climate regimes. Such localized climatic behaviour is expected to induce a negative association with the surrounding, and this is confirmed by Figure \ref{fig:IMD_DTR_SVD12s_res_corr_4cities.pdf} for cities like Delhi, Mumbai and Chennai.

\noindent\textbf{Orographic effect on temperature:} The Western Ghats force moisture-laden winds to rise, resulting in heavy rainfall (orographic effect) on the windward side and dry, hotter conditions in the leeward rain-shadow regions. This could be an additional source of the disparate association one observes between Mumbai, which is to the west of the Western Ghats and regions towards east of the same mountain range (see Figure \ref{fig:IMD_DTR_SVD12s_res_corr_4cities.pdf}). 

\noindent\textbf{Spatial asymmetry in association:} Our investigation brings out further features of differing spatial association, with respect to a fixed location, in the directions of latitude and longitude. More generally, fix a grid $g_1$, and consider the grid $g_2$ that has the highest correlation with $g_1$. Consider the difference in their spatial locations in latitude or longitude. The distribution of the difference in latitude across all grids has a significant mode at $0$, while the same 
in longitudes is approximately uniform around $\pm 1$. We have presented these two distributions in \cite{bhatbosesupport}.
Such salient spatial patterns were not visible earlier, i.e. without trimming the temporal influence. 
\vskip5pt

\subsubsection{Spatio‑climatological correlation matrix} 
\begin{figure}[h]
    \centering
    \includegraphics[width=0.3\linewidth]{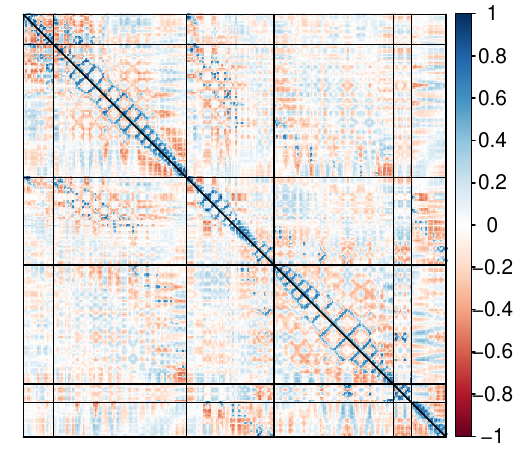}
    \caption{Correlations with grids arranged first according to climatic regions and then in spiral Hilbert space filling curve manner. \textit{Upper triangle},  Pearson correlation matrix ($R^S$) based on trimmed data ($S$); \textit{Lower triangle}, the MP de-noised version of it.}
    \label{fig:Spiral_ord_clmtc_reg_corr_mat_with_MPdns}
\end{figure}
 Figure \ref{fig:Spiral_ord_clmtc_reg_corr_mat_with_MPdns} presents the correlation matrix $R^S$ with the grids ordered, first according to climatic regions and then in spiral Hilbert space filling curve manner. This enables clearer visualization of spatial association patterns within each climatic region, and proximity within the matrix implicitly indicates actual spatial proximity of the locations.

Organizing the locations according to climatic regions is justified intuitively, and is also 
substantiated by mathematical investigation. For example, analysis of the eigenvectors corresponding to significant eigenvalues as per the MP law supports the observation of climatic region-specific spatial association pattern (details are not reported here).
\vskip5pt
\newpage

\noindent\textbf{Temporal changes in spatial association:} 
The inherent assumption in obtaining the preceding pattern is that it is temporally static, and hence we used the entire data. It is also possible to investigate such patterns at a smaller time interval, such as a calendar year, or even a calendar month. 
Carrying out such an investigation would require studying numerous such matrices. A mathematically compact manner of summarizing these matrices would be to consider the ESD. Note that 
the correlation matrices and the ESDs offer $280\times 280$ and $280\times 1$ dimensional summaries respectively. 

\noindent \textbf{(i) Capturing dependence as linear association:} We obtained the ESD of the correlation matrices $\{R^{S}_i\}$ (see \cite{bhatbosesupport}),
and the studies clearly indicate the occurrence of a drastic change in spatial association in the late 1960s. 

\vskip5pt


\noindent \textbf{(ii) Capturing general statistical dependence:}  
\begin{figure}[h]
\centering
  \begin{subfigure}{.3\textwidth}
      \subfloat[a][Yearly all India data]{\includegraphics[width=1\linewidth]{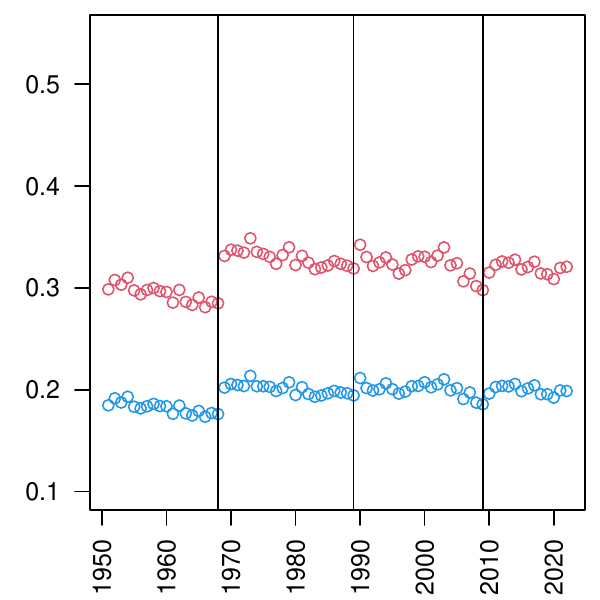}
      \label{fig:SB_yrly_SVD10s_DTR}} \\
      \subfloat[b][Monthly all India data]{\includegraphics[width=1\linewidth]{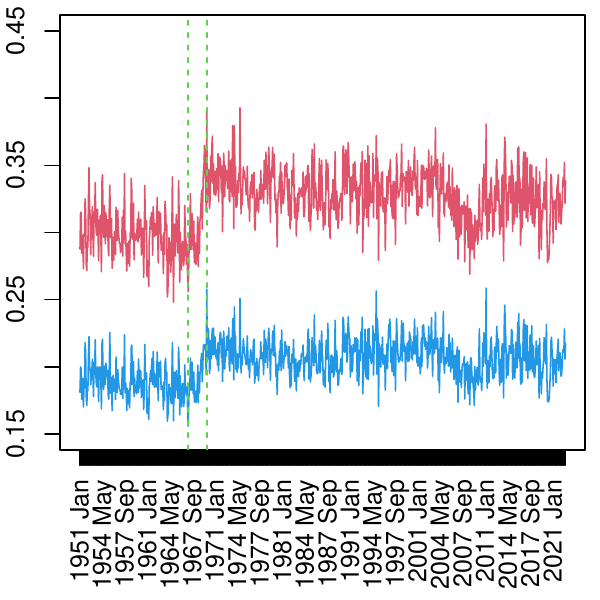}
      \label{fig:DTR_SVD12s_mnly_SB_W1_W2.pdf}}
  \end{subfigure}
  \begin{subfigure}{.6\textwidth}
  \centering
    \includegraphics[width=1\linewidth, height=3.5in]{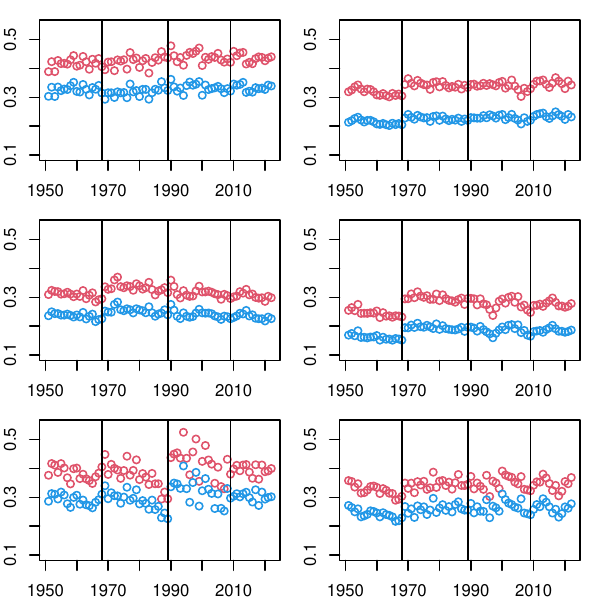}
    \caption{Yearly climatic region level data}
    \label{fig:SVD_yrly_clmtc_reg_SVD10s_DTR}
  \end{subfigure}
\caption{Spatial Bergsma statistics ($S_B$) at various spatio-temporal resolution based on trimmed data ($S$) and 
for lag-1 adjacency (\textit{red}),  and exponential distance decay (\textit{blue}).}
\label{fig:SB_stat}
\end{figure}
The Spatial Bergsma statistic $S_B$ provides a univariate numerical summary, and a small value indicates statistical independence. We computed $S_B$ based on $S$ at temporal resolution of, (i) year, (ii) month within each year, and at spatial resolution of, (i) all-India, (ii) six climatic regions.  The results, presented graphically in Figure \ref{fig:SB_stat}, show that the degree and variation in association are climatic region specific. Further, it shows without a doubt that there has been at least one, and possibly more,  major change(s) in spatial association pattern over the Indian region with the 72 years considered here. These have had varying degrees of effect in different climatic regions. We are able to confirm the occurrence of a drastic change in the spatial association pattern in the late 1960's by the use of $S_B$-statistic based on $S$ at the various spatial and temporal resolutions mentioned above.
\vskip5pt

\subsubsection{Climatic teleconnection detection}

Climatic teleconnection detection involves identifying spatially correlated, large-scale pressure and circulation anomalies (e.g., ENSO, NAO) that influence distant weather patterns. Climate teleconnections, primarily ENSO and Indian Ocean SSTs, significantly influence India's temperature variability.

\begin{figure}
\centering
\begin{subfigure}{.3\textwidth}
  \centering
    \includegraphics[width=1.8in, height=1.8in]{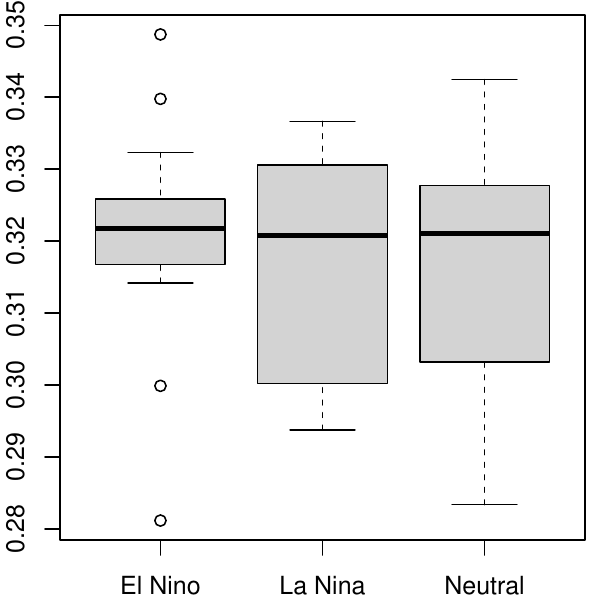}
    \caption{ENSO-SB distribution}
  \label{fig:ENSO_SB_lag1_SVD10s_DTR}
\end{subfigure}
\begin{subfigure}{.6\textwidth}
  \centering
    \includegraphics[width=3.8in, height=2.2in]{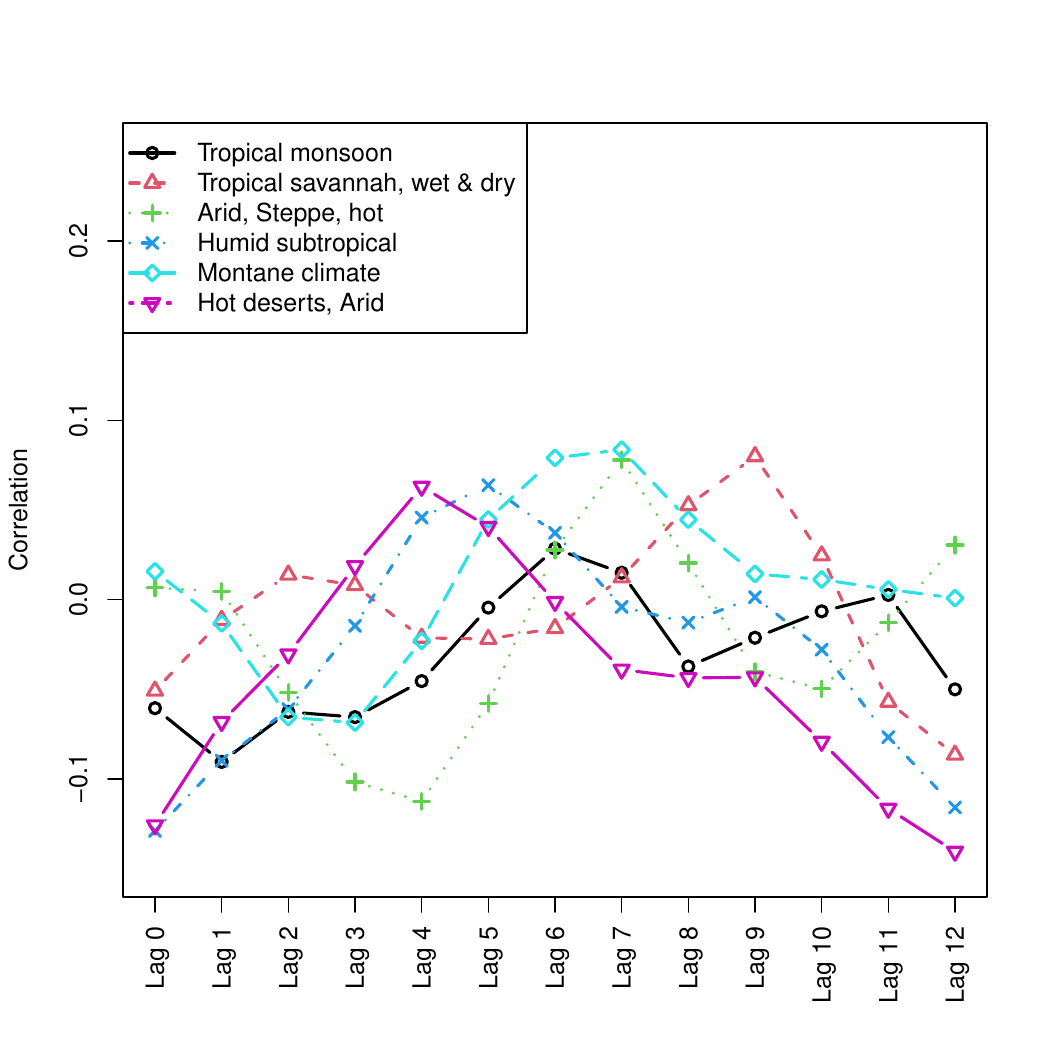}
  \caption{IOD-SB correlation}
  \label{fig:SST_mnthly_clmt_reg_SB_corr}
\end{subfigure}
\caption{Teleconnection and core spatial association (a) Categorized by ENSO information, distribution of Spatial Bergsma statistics ($S_B$) from individual year's trimmed data ($S$); (b) Correlations (instantaneous and lagged) between monthly SST of Indian Ocean and monthly Spatial Bergsma statistics ($S_B$), at climatic region levels for India.}
\label{fig:teleonnect}
\end{figure}

\noindent\textbf{ENSO and core spatial association:} El Ni\~no Southern Oscillation (ENSO) is known to be strongly linked to Indian Summer Monsoon Rainfall, with El Ni\~no conditions often reducing rainfall. However, it also influences extreme temperature events (warm/cold days/nights). We investigated the impact of ENSO on the distributional behavior of yearly $S_B$. Figures  \ref{fig:Spiral_ord_clmtc_reg_corr_mat_with_MPdns} and \ref{fig:SB_stat} have  established that climatic regions have different impacts on centrality, and  dispersion of $S_B$. We \textit{augmented} the annual $S_B$ statistics values with the ENSO data. Figure \ref{fig:teleonnect} panel (a) shows the impact of ENSO on the distributional behavior of $S_B$. The variation in $S_B$ is less in the El Ni\~no phase, possibly because the influence of the phase is so strong that it takes over and dictates the variation in DTR. 

\noindent\textbf{Indian ocean IOD and core spatial association:} Variability in sea surface temperatures (SST) over the Indian Ocean exerts a major influence on maximum temperatures across India. A key driver of this variability is the Indian Ocean Dipole (IOD), a climate pattern characterized by shifts in oceanic heat distribution. During a positive IOD phase, warm waters accumulate in the western Indian Ocean while cold, deep waters upwell in the eastern basin; the reverse occurs during a negative phase. These phases typically persist from several weeks to many months, making the IOD a seasonal climate index of considerable importance. DMI is a numerical index, usually calculated as the difference in SST anomalies between the western Indian Ocean and the southeastern Indian Ocean. We compared the monthly $S_B$ statistics presented in Figure \ref{fig:SB_stat}(b) and (c) with the monthly DMI values for the Indian Ocean (obtained from \cite{tokyodipole}). Our findings suggest that increases in SST are associated with weakening of statistical dependence between grids within a climatic region (see Figure \ref{fig:teleonnect} panel (b)).


\subsubsection{Validation}

\begin{figure}[h]
\centering
\begin{subfigure}{.3\textwidth}
  \centering
    \includegraphics[width=1.8in, height=1.8in]{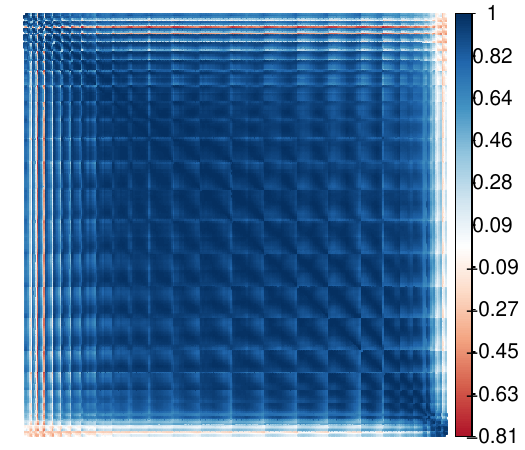}
    \caption{Original DTR data $D$.}
  \label{fig:CRU_cor_MPdns}
\end{subfigure}
\begin{subfigure}{.3\textwidth}
  \centering
    \includegraphics[width=1.8in, height=1.8in]{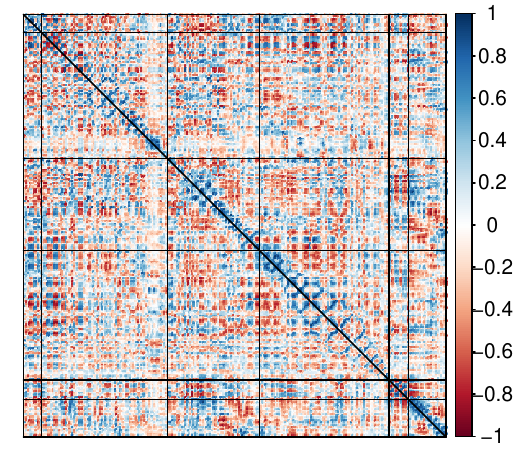}
  \caption{Trimmed DTR data $S$.}
  \label{CRU_SVD12s_clmt_cor_MPdns}
\end{subfigure}
\caption{Correlation matrices 
based on monthly DTR data of India from CRU, \\
(a)
 based on original data; (b)
 based on trimmed data with 12 SVs and grids arranged according to climatic regions.}
\label{fig:cru-cor}
\end{figure}

\noindent\textbf{Validation with alternate Indian data}:\label{sec:alt-IN}
We used our analysis pipeline on a \textit{monthly} data for the same 72 years and $1^{\circ}$ resolution from \cite{cru} (url: https://crudata.uea.ac.uk/cru/data/hrg/ version \verb|cru_ts4.07|). 
We observe the same spatial pattern from the monthly DTR data as was identified by the daily DTR data (compare Figures \ref{fig:DTR_cor_MPdns} and \ref{fig:Spiral_ord_clmtc_reg_corr_mat_with_MPdns} with Figure \ref{fig:cru-cor}). 

\begin{figure}[ht!]
\centering\includegraphics[width=0.6\linewidth, height=1.5in]{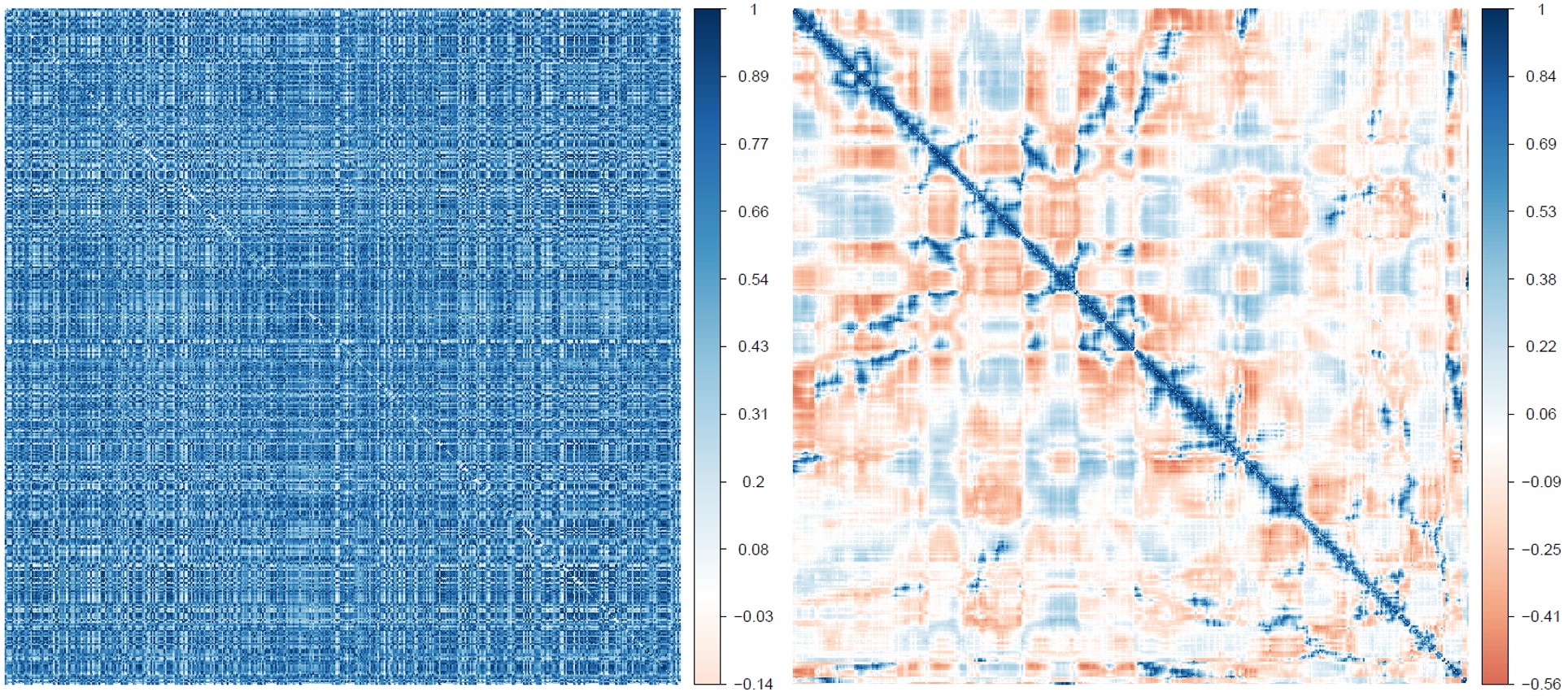}
   \caption{Correlation matrices for DTR from 417 municipalities of Bahia: 
   original data (left); trimmed DTR with 5 SVs, municipalities arranged by spatial proximity (right).}
\label{fig:bahia-cor}
\end{figure}

\noindent\textbf{Bahia data with differing spatio-temporal resolution:}
We also looked briefly at daily DTR data of Bahia, Brazil, from \cite{cidacs}.
This is a non-gridded data, at a different spatial resolution, for a smaller area, with higher number of time series, over a climatologically limited area for only 23 years. We hope to report a full analysis soon elsewhere, but Figure \ref{fig:bahia-cor} already shows a novel spatial pattern.


\section{Conclusions}\label{sec:conclusions} 
This study advances the methodological foundations of spatial time series analysis by introducing a Random Matrix Theory (RMT)–based pipeline to isolate core spatial associations in climate data. By systematically detrending temporal components we demonstrate how subtle yet significant spatial signals can be recovered from data otherwise dominated by strong temporal co‑evolution. The RMT based  approaches enables detrending, denoising of high‑dimensional correlation matrices and also provided robust low‑dimensional summaries that facilitated comparison of such matrices across regions and time windows.

Applied to the diurnal temperature range (DTR) data of India, this framework revealed distinct spatial anomalies in spatial association shaped by topography, mesoclimate, and urbanization. Centrality and dispersion of the spatial association measures are impacted by the climatic regions. We also found that there is a strong association in the immediate vicinity for each of the spatial locations, however, with an asymmetric pattern slightly differing in north-south and east-west directions. Spatial associations were also seen to be affected by global climatic phenomena like ENSO (especially by the El Ni\~no phase) and regional phenomena like IOD. Influence of ENSO has been reported earlier, for example, by \cite{vinnarasi2019}. Importantly, this method highlighted how spatial dependence evolves over time, offering insights into the propagation of climate anomalies and the clustering of risks across regions. The yearly ESD analysis clearly indicated a drastic change in the spatial association pattern in the late 1960s. $S_B$-statistics confirmed that there has been at least one major temporal change (around 1968-69) in the spatial association pattern in DTR, over India.

Beyond the Indian application, the proposed methodology is broadly applicable to diverse spatio‑temporal datasets, including those from other climatic regions. From theoretical perspective, distributional results on the GSVs of random data matrices appear to be absent from the literature and is worth pursuing. The core spatial association/dependence pattern that emerged in our analysis could help build composite STS models for DTR data, which is otherwise rather difficult due to the strong non-isotropic and non-separable space-time dependence.

By bridging statistical innovation with climate science, this work contributes to both theoretical advancement and practical utility. The core spatial association matrix, obtained in a black-box method, has the potential to discover yet unknown climate anomalies, without the requirement of explcit complex modelling, e.g. identification of “hot spots” and “cold spots,” as well as the disruption of smooth spatial behaviour.

The ability to disentangle temporal and spatial signals enhances predictive modelling, supports resilience planning, and strengthens the evidence base for policy interventions in the face of accelerating climate change. Future research may extend this framework to compound hazards, multi‑variable associations, and integration with dynamical climate models, thereby deepening our understanding of spatial dependence in complex environmental systems.
\vskip7pt



\noindent \textbf{Supporting information, \cite{bhatbosesupport}}. 
This document contains a brief discussion on the theory behind SVD, GSVD, spectral behaviour of eigen-values and singular-values. It also includes additional information on the analysis of the Indian DTR data, particularly with respect to: (a) preliminary exploration,  (b) trimming and GSVD, (c) effect of trimming on ESDs, (c) location-specific spatial association, and (d) temporal changes in spatial association matrix. 
\vskip7pt

\noindent \textbf{Data availability statement}. 
The data used in this study were derived from the following resources available in the public domain:\vskip3pt

\noindent (i) Climate Research Services, India Meteorological Department, Guided Data Archive, at \url{https://imdpune.gov.in/lrfindex.php}.
\vskip3pt
 
\noindent (ii)  Climate Research Unit, East  Anglia  University, UK, at\\ https://crudata.uea.ac.uk/cru/data/hrg/ version \verb|cru_ts4.07|.
\vskip3pt

\noindent (iii) Center for Data and Knowledge Integration for Health (CIDACS), Bahia, Brazil, at \verb|https://cidacs.bahia.fiocruz.br/english_/|
\vskip3pt

\noindent (iv) Tokyo Climate Center, at\\  
\verb|https://ds.data.jma.go.jp/tcc/tcc/products/elnino/index/iod_index.htm|
\vskip7pt

\noindent \textbf{Funding statement}. The authors acknowledge the support from 
(i) J.C. Bose National Fellowship, JBR/2023/000023 from Anusandhan National Research Foundation,  Govt. of India, and 
(ii) Dame Kathleen Ollerenshaw (DKO) Research Visitor award from the University of Manchester, UK, June 2025.

\noindent \textbf{Conflict of interest}. The authors declare that they have no conflict of interest. 


\bibliography{dtr_ref1}

\end{document}